\documentclass[prb,amsmath,amssymb,superscriptaddress,twocolumn]{revtex4}
\usepackage{float}
\usepackage{amsmath}
\usepackage{amssymb}
\usepackage{amsfonts}
\usepackage{euscript}
\usepackage{enumerate}
\usepackage{hhline}
\usepackage{pslatex}
\usepackage{tabularx}
\usepackage[usenames,dvipsnames]{xcolor}

\usepackage{graphicx}

\usepackage{dcolumn}
\usepackage{bm}
\usepackage[sort&compress]{natbib}

\usepackage{lipsum}
\usepackage{textcomp, gensymb}

\makeatletter
\renewcommand{\@biblabel}[1]{#1. }
\renewcommand{\@dotsep}{500}
\renewcommand{\@pnumwidth}{0em}
\renewcommand{\l@figure}[2]{
\@dottedtocline{1}{1.5em}{2em}{Figure #1}{}\vspace{15pt}}

\usepackage[normalem]{ulem}
\usepackage{upgreek}

\begin{document}

\title{Advancing on-chip Kerr optical parametric oscillation towards coherent applications covering the green gap}

\author{Yi Sun}
\thanks{These two authors contributed equally}
\affiliation{Microsystems and Nanotechnology Division, Physical Measurement Laboratory, National Institute of Standards and Technology, Gaithersburg, MD 20899, USA}
\affiliation{Joint Quantum Institute, NIST/University of Maryland, College Park, MD 20742, USA}

\author{Jordan Stone}
\thanks{These two authors contributed equally}
\affiliation{Microsystems and Nanotechnology Division, Physical Measurement Laboratory, National Institute of Standards and Technology, Gaithersburg, MD 20899, USA}
\affiliation{Joint Quantum Institute, NIST/University of Maryland, College Park, MD 20742, USA}

\author{Xiyuan Lu}\email{xnl9@umd.edu}
\affiliation{Microsystems and Nanotechnology Division, Physical Measurement Laboratory, National Institute of Standards and Technology, Gaithersburg, MD 20899, USA}
\affiliation{Joint Quantum Institute, NIST/University of Maryland, College Park, MD 20742, USA}

\author{Feng Zhou}
\affiliation{Microsystems and Nanotechnology Division, Physical Measurement Laboratory, National Institute of Standards and Technology, Gaithersburg, MD 20899, USA}
\affiliation{Joint Quantum Institute, NIST/University of Maryland, College Park, MD 20742, USA}

\author{Zhimin Shi}
\affiliation{Reality Labs Research, Meta, Redmond, Washington 98052, USA}

\author{Kartik Srinivasan}\email{kartik.srinivasan@nist.gov}
\affiliation{Microsystems and Nanotechnology Division, Physical Measurement Laboratory, National Institute of Standards and Technology, Gaithersburg, MD 20899, USA}
\affiliation{Joint Quantum Institute, NIST/University of Maryland, College Park, MD 20742, USA}

\date{\today}

\begin{abstract}
     \noindent Optical parametric oscillation (OPO) in Kerr microresonators can efficiently transfer near-infrared laser light into the visible spectrum. To date, however, chromatic dispersion has mostly limited output wavelengths to $>560$~nm, and robust access to the whole green light spectrum has not been demonstrated. In fact, wavelengths between $532$~nm and $633$~nm, commonly referred to as the ``green gap'', are especially challenging to produce with conventional laser gain. Hence, there is motivation to extend the Kerr OPO wavelength range and develop reliable device designs. Here, we experimentally show how to robustly access the entire green gap with Kerr OPO in silicon nitride microrings pumped near $780$~nm. Our microring geometries are optimized for green-gap emission; in particular, we introduce a dispersion engineering technique, based on partially undercutting the microring, which not only expands wavelength access but also proves robust to variations in resonator dimensions, in particular, the microring width. Using just two devices, we generate $>100$ wavelengths evenly distributed throughout the green gap, as predicted by our dispersion simulations. Moreover, we establish the usefulness of Kerr OPO to coherent applications by demonstrating continuous frequency tuning ($>50$~GHz) and narrow optical linewidths ($<1$~MHz). Our work represents an important step in the quest to bring nonlinear nanophotonics and its advantages to the visible spectrum.
\end{abstract}

\maketitle
\section{Introduction}
\begin{figure*}[t!]
\centering\includegraphics[width=0.90\linewidth]{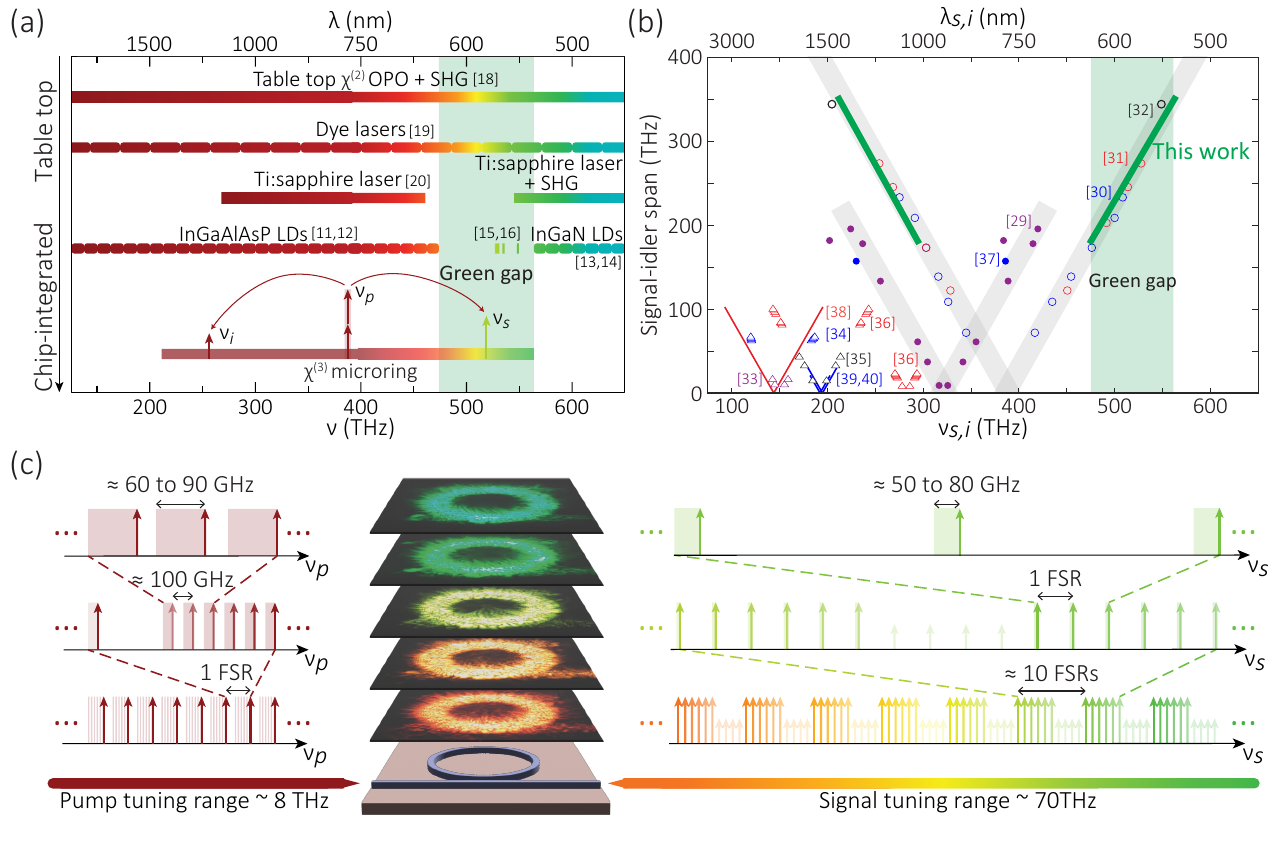}
\caption{\textbf{Aiming for the green gap with Kerr microresonator optical parametric oscillation (OPO).} \textbf{a} Comparison of currently available laser technologies. Table-top OPO and second harmonic generation (SHG) provide broad spectral coverage using bulk free-space optics \cite{Sperling_RSI_2021}. Commercial dye lasers, which are also table-top technologies, rely on potentially toxic chemicals and face a limitation in that each dye only offers a tuning range of $50$~nm to $100$~nm, necessitating multiple dyes to achieve a broader range of tuning capabilities \cite{schafer_book_2013}. Ti:sapphire lasers are highly tunable and have recently been demonstrated on-chip \cite{Wang_NatPhoton_2023}, but they exhibit (even when combined with SHG) a large spectral gap for orange, yellow, and green colors. Commercially-available laser diodes can emit a wide range of infrared and visible wavelengths, but the active region (which is composed in tunable ratios of two materials) emission spectrum exhibits a gap between $532$~nm and $633$~nm \cite{Hamada_IEEEJQE_1993,Kitatani_JJAP_2000,Sun_NatPhoton_2016, Ra_SciAdv_2020} with few exceptions \cite{Mei_LSA_2017,Li_APL_2022}. This gap has become known as the ``green gap''. On the other hand, OPO in Kerr microresonators can generate orange, yellow, and green colors without spectral gaps. \textbf{b} Comparison of chip-integrated OPO results, charting demonstrated signal and idler frequencies ($\nu_{\rm{s}}$ and $\nu_{\rm{i}}$) and their separation for various $\chi^{(2)}$ and $\chi^{(3)}$ (Kerr) systems. While $\chi^{(2)}$ OPOs are mostly focusing on infrared wavelengths above $1400$~nm shown in red and blue lines \cite{Ledezma_SciAdv_2023,Lu_Optica_2021,Bruch_Optica_2019}, Kerr OPOs could span from infrared to the visible wavelengths, as indicated by the triangles and circles \cite{Lu_Optica_2019C, Stone_APLPhoton_2022,Lu_Optica_2020, Domeneguetti_Optica_2021, Lu_OL_2022,Tang_OL_2020,pidgayko_Optica_2023,Black_Optica_2022,Perez_NatCommun_2023}. However, a comprehensive coverage of green gap region has never been demonstrated. \textbf{c} Illustration of three pump-actuated OPO tuning mechanisms. When the pump laser (frequency $\nu_{\rm{p}}$) is switched between adjacent longitudinal modes, i.e., $\nu_{\rm{p}}$ shifts by one free spectral range (FSR), $\approx900$ GHz, $\nu_{\rm{s}}$ changes by $\approx9$~THz (bottom row). When $\nu_{\rm{p}}$ is tuned within one cavity mode, thermo-optic effects induce changes to dispersion that cause $\nu_{\rm{s}}$ to mode hop in FSR increments (middle row). In between mode hops, $\nu_{\rm{s}}$ tunes continuously with $\nu_{\rm{p}}$ with an approximately $6$:$5$ ratio (top row), resulting in up to $80$~GHz of tuning. The stacked color images are of scattered light for different signal wavelengths generated by the $\mu$OPOs in this work.}
\label{Fig1}
\end{figure*}

\noindent The development of compact visible lasers will benefit numerous sectors of science and industry, including laser lighting and displays \cite{Chellappan_AO_2010, Zhao_NatCommun_2019}, spectroscopy for timekeeping and sensing \cite{Hummon_Optica_2018, Shang_OE_2020, zou2019sensing}, medical practices \cite{Luke_JofLasersinMedicalSci_2019}, and quantum technology \cite{Moody_JoPhysPhoton_2022, Uppu_NatNano_2021}. While progress has been made in the blue and red wavelength regions, a lack of efficient and compact green laser sources, also known as the ``green gap'' problem (see Fig.~\ref{Fig1}a), still plagues the laser market \cite{Pleasant_NatPhoton_2013,Moustakas_RPP_2017}. III-V semiconductor lasers provide a compelling combination of efficiency and small size \cite{Hamada_IEEEJQE_1993,Kitatani_JJAP_2000,Sun_NatPhoton_2016, Ra_SciAdv_2020}, but they require Watts of input power and often (especially at “green gap” wavelengths) lack the spectral purity needed for high-coherence applications \cite{Mei_LSA_2017,Li_APL_2022}. Injection locking Fabry-Perot diode lasers to high-finesse microresonators can improve coherence, but the output wavelengths are constrained by the availability of pump lasers and, so far, are continuously tunable over only a few GHz \cite{Corato_NatPhoton_2022}. In Fig.~\ref{Fig1}a, we compare various commercial solutions to the green gap problem \cite{Sperling_RSI_2021, schafer_book_2013, Wang_NatPhoton_2023}, charting them by their size and wavelength range.

Another way to produce green laser light is through nonlinear optical processes. This is the strategy adopted most by industry, and it offers an intriguing path to scalability via photonic integration since small optical volumes promote efficient nonlinear interactions (commercial instruments using bulk optical components are typically $\approx~1$~m$^3$ in size). For example, nonlinear microresonators can generate the frequency harmonics of near-infrared pump lasers to produce visible light, albeit with limited wavelength tuning capability \cite{Carmon_NatPhys_2007,Ling_LaserPhotonicsReviews_2023, Guo_Optica_2016a, Surya_Optica_2018, Levy_OE_2011, Lu_NatPhoto_2021, Nitiss_NatPhoton_2022}. Alternatively, widely-separated Kerr optical parametric oscillation (OPO) is a flexible approach to generate visible light by four-wave mixing (FWM) from, e.g., a near-infrared pump, and in recent years, OPO based on FWM in optical microresonators (we dub such devices ``$\mu$OPOs'') has been investigated \cite{Sayson_NatPhoton_2019, Lu_Optica_2019C, Stone_APLPhoton_2022,Lu_Optica_2020, Domeneguetti_Optica_2021, Lu_OL_2022,Tang_OL_2020,pidgayko_Optica_2023,Black_Optica_2022,Perez_NatCommun_2023}. In these systems, energy from a monochromatic pump laser with frequency $\nu_{\rm{p}}$ is transferred to a blue-shifted signal wave ($\nu_{\rm{s}}$) and red-shifted idler wave ($\nu_{\rm{i}}$), as shown in Fig.~\ref{Fig1}a. Visible $\mu$OPOs can operate with milliwatt-level threshold powers and have shown pump-to-sideband conversion efficiencies up to 15~\% \cite{Lu_Optica_2019C, Stone_APLPhoton_2022}. In Fig.~\ref{Fig1}b, we compare the operating wavelengths and spectral separations, $\nu_{\rm{s}}$-$\nu_{\rm{i}}$, reported in several $\mu$OPO studies, including, for completeness, several demonstrations of OPO in $\chi^{(2)}$ nanophotonics \cite{Ledezma_SciAdv_2023,Lu_Optica_2021,Bruch_Optica_2019}. Importantly, signal frequencies in the green spectrum have been reported~\cite{Stone_APLPhoton_2022,Lu_Optica_2020, Domeneguetti_Optica_2021}, but the highest frequency reported so far is $\approx548.9$~THz~\cite{Domeneguetti_Optica_2021}, which is $\approx14.6$~THz shy from the edge of the green gap. In addition, the $\mu$OPO output power and wavelength are sensitive to external parameters like temperature, pump power, and pump-resonator detuning \cite{Stone_PRA_2022}, as well as to microring geometry. These sensitivities tend to grow in proportion to the $\mu$OPO separation ($\nu_{\rm{s}}-\nu_{\rm{i}}$) [Ref.~\onlinecite{Lu_Optica_2020}] and therefore present a major challenge for $\mu$OPOs aiming at more comprehensive coverage of the green gap.  

\begin{figure*}[t!]
\centering\includegraphics[width=0.90\linewidth]{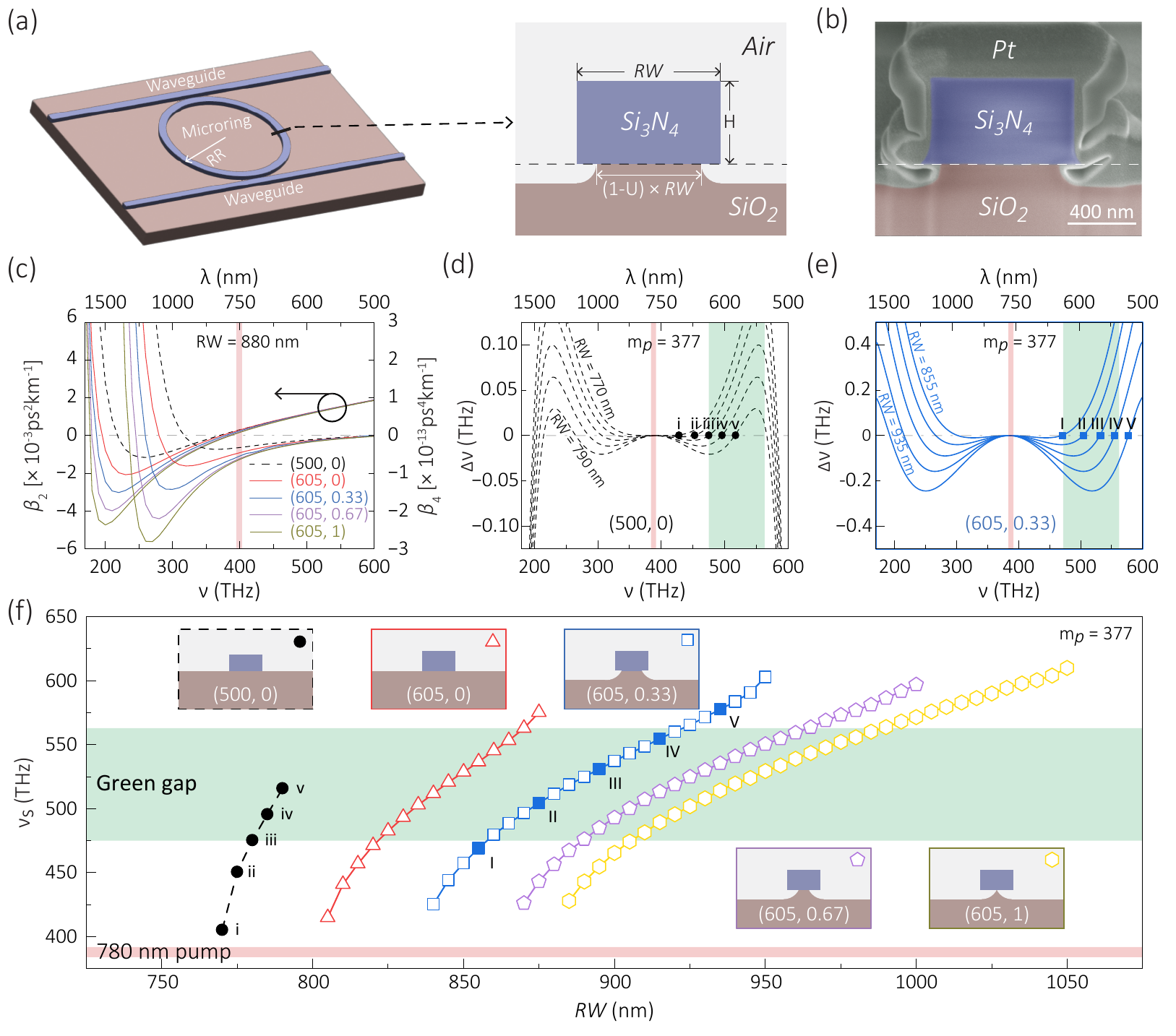}
\caption{\textbf{Microresonator dispersion design.} \textbf{a} Left: Illustration of a microring OPO device, indicating the outer ring radius ($RR$) and two coupling waveguides that are used for separately extracting the signal and idler waves. Right: Illustration of the microring cross section. The device has a silicon nitride (Si$_3$N$_4$, hereafter SiN) core, a silicon dioxide (SiO$_2$) substrate, and top air cladding. Its dimensions are defined as ring width ($RW$) and height ($H$). We also use potassium hydroxide (KOH) to etch the SiO$_2$ substrate underneath the SiN core for dispersion engineering. The etch is quantified by the undercut parameter, $U$. \textbf{b} False-color cross-sectional scanning electron microscope image from one device milled by focused ion beam. \textbf{c} Simulations of the second- and fourth-order expansion coefficients ($\beta_{\rm{2}}$ and $\beta_{\rm{4}}$) of the TE0 mode propagation constant for different microring geometries (parameterized as ($H, U$) and $RW$ = 880~nm). \textbf{d} Simulated frequency mismatch spectra for five devices with ($500$, 0) and $RW$ ranging from 770 nm to 790 nm with increments of 5 nm. Points i to v marked with solid circles indicate the zero crossings, which predict the signal frequency ($\nu_{\rm{s}}$). \textbf{e} Simulated frequency mismatch spectra for five devices with ($605$, 0.33) and $RW$ ranging from 855 nm to 935 nm with increments of 20 nm. Points I to V marked with solid squares indicate the zero crossings, which predict the $\nu_{\rm{s}}$. \textbf{f} Simulated values of $\nu_{\rm{s}}$ versus $RW$. For \textbf{d}, \textbf{e} and \textbf{f}, the pump mode azimuthal number, $m_{\rm{p}}$, is fixed at $377$. In general, thicker SiN increases higher-order dispersion to broaden the OPO spectral coverage. We find $H=605$ nm is sufficient to cover the green gap. Furthermore, increasing $U$ not only increases spectral coverage but also decreases the sensitivity of $\nu_{\rm{s}}$ to $RW$, resulting in more robust green light generation and deeper control (e.g., via the tuning mechanisms described in Fig. \ref{Fig1}) over the OPO spectrum. }
\label{Fig2}
\end{figure*}

Here, we use $\mu$OPOs to access the entire green gap, achieving the highest frequency of $\approx563.51$~THz, increasing wavelength access by $\approx14.2$~nm beyond the previous record in Ref.~\onlinecite{Domeneguetti_Optica_2021} and improving robustness with respect to parameter variations. Using just two devices, we can selectively generate $>100$ $\mu$OPOs, each with a unique green-gap signal frequency that is separated from its nearest neighbor by roughly the microresonator free spectral range (FSR). This breakthrough is enabled by a novel dispersion design in which the substrate is partially etched away, so that a greater portion of the microresonator is air-clad. We perform simulations and measurements to explore the effects of such an undercut on the $\mu$OPO. In particular, increasing the undercut makes $\nu_{\rm{s,i}}$ less sensitive to $\nu_{\rm{p}}$ and device dimensions; hence, one device supports many green-gap $\mu$OPOs, and spectral gaps are filled in using a second device with different dimensions. Our ability to generate a multitude of $\mu$OPOs within a single device stems from unique tuning mechanisms (two mode hop-based processes for coarse tuning and one process for continuous fine tuning) that we depict in Fig. \ref{Fig1}c. Finally, to prove our $\mu$OPOs are well-suited to coherent applications in the green gap, we present measurements of heterodyne beatnotes between the $\mu$OPO signal and a separate narrow-linewidth laser, and we characterize the $\mu$OPO continuous frequency tunability. We measure fitted linewidths below $1$~MHz and continuous tuning ranges up to $80$~GHz. With further integration, including recent advancements in chip-integrated 780~nm lasers \cite{Isichenko_arXiv_2023,Corato_NatPhoton_2022,Zhang_Optica_2023}, $\mu$OPOs are a realistic solution to the green gap problem, especially when low noise is required. 

\section{Results}
\noindent \textbf{Device design.}
In Figure 2, we present images of a nominal microring device as well as simulations of the chromatic dispersion. We fabricate microrings out of stoichiometric silicon nitride (Si$_3$N$_4$, hereafter written as SiN) with outer ring radius $RR=25$~$\mu$m and nominal height $H=605$~nm; the microrings sit on a SiO$_2$ lower cladding and are air-clad on the sides and top (See Materials and Methods for details). We couple light in/out of the microrings via two bus waveguides, as shown in Fig. \ref{Fig2}a. One waveguide is narrower and runs closer to the microring; it in/out-couples pump and signal light. The other waveguide is wider and farther from the microring and is used to out-couple the long-wavelength idler, which cannot propagate in the narrower waveguide (the waveguide is cut-off at the idler wavelength). Using heated potassium hydroxide, we undercut the microrings by an amount $U$ that can be between $0$ (no undercut) and $1$ (completely undercut). In Fig.~\ref{Fig2}b, we show a scanning electron microscope image of the microring cross-section in which the undercut ($U \approx0.25$) is clearly visible. 

We choose $H$ and $U$ to optimize dispersion, which we parameterize using the frequency mismatch, $\Delta \nu = \nu_{\mu}+\nu_{-\mu}-2\nu_{\rm{0}}$, where $\nu_{\mu}$ is the frequency of a mode whose longitudinal mode number (with respect to the pump mode) is $\mu$. In general, mode pairs with small positive $\Delta \nu$ can oscillate \cite{Lu_Optica_2020,Lu_Optica_2019C,Stone_APLPhoton_2022}; hence, to realize $\mu$OPOs with wide frequency separations, we desire strong normal group velocity dispersion (GVD) in the pump band (negative curvature of $\Delta \nu$ around $\nu_{\rm{0}}$) and higher-order GVD to balance $\Delta \nu$ away from $\nu_{\rm{0}}$. To understand the relationships between $H$, $U$, and GVD, in Fig. \ref{Fig2}c we present the simulated GVD coefficients, $\beta_{\rm{2}}$ and $\beta_{\rm{4}}$, for the fundamental transverse electric polarized (TE0) modes of five ring resonators with different ($H$, $U$) values. In the pump band, $\beta_{\rm{2}}$ is slightly positive (indicating normal dispersion) and nearly independent of $H$ and $U$, while $\beta_{\rm{4}}$ becomes significantly more negative for increasing $H$ and $U$ (indicating greater higher-order dispersion that can balance the normal dispersion for mode pairs far from the pump). Moreover, we can make the general observation that the idler band (where $\beta_{\rm{2}}$ and $\beta_{\rm{4}}$ are more sensitive to $\nu$, $H$, and $U$) primarily determines the geometric dispersion. Alternatively, we can study the $\Delta \nu$ spectrum and its dependence on $H$ and $U$. In Figs. \ref{Fig2}d and \ref{Fig2}e, we present five $\Delta \nu$ spectra each for devices with ($500$, $0$) and ($605$, $0.33$), respectively, with systematic variations to $RW$. In the $\Delta \nu$ space, zero crossings (marked in Figs. \ref{Fig2}d and \ref{Fig2}e by solid circles and squares, respectively) determine $\nu_{\rm{s}}$, so we can predict specifically the dependence of $\nu_{\rm{s}}$ on $RW$, $H$, and $U$. In Fig. \ref{Fig2}f, we plot $\nu_{\rm{s}}$ versus $RW$ for several ($H, U$) pairs. We find that in the ($500$, $0$) configuration that has been extensively used in prior studies~\cite{Lu_Optica_2020,Lu_Optica_2019C,Stone_APLPhoton_2022}, $\nu_{\rm{s}}$ is limited to less than $530$~THz, and a relatively narrow range of $RW$ values allow for green gap emission. However, increasing $H$ and $U$ has two notable effects: The maximum realizable $\nu_{\rm{s}}$ is increased, and the $\mu$OPO is more robust to geometry perturbations. Specifically, the slope $d\nu_{\rm{s}}/dRW$, which quantifies the $\mu$OPO sensitivity to device dimensions, decreases for increasing $U$. As a result, more green-gap $\mu$OPOs become supported by a given device, and designs are more tolerant to fabrication uncertainties; in the future, it may be possible to generate cyan or even blue signal colors. In the next section, we experimentally verify these concepts and leverage them towards comprehensive access of the green gap.

\begin{figure*}[t!]
\centering\includegraphics[width=0.90\linewidth]{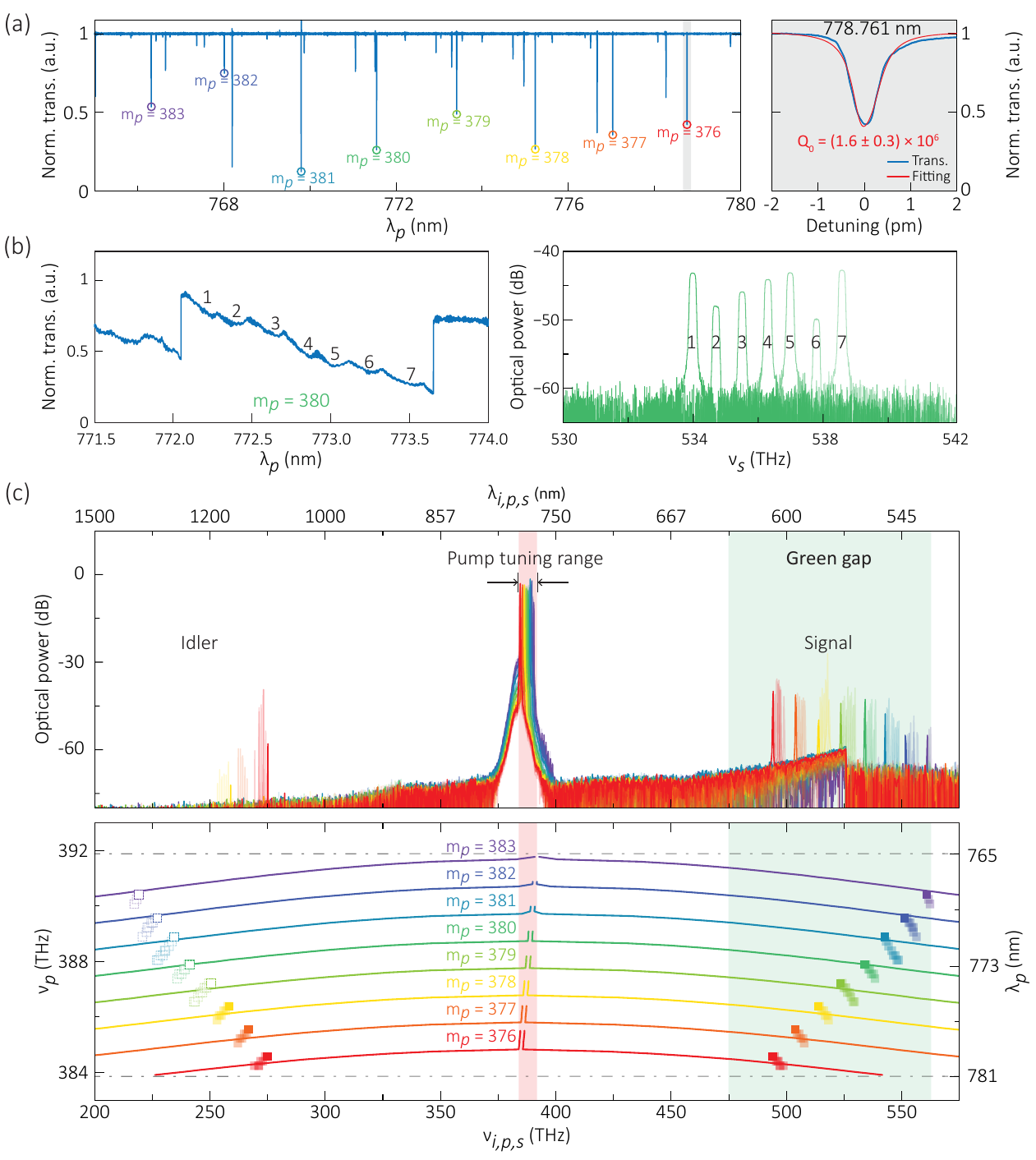}
\caption{\textbf{Green gap OPO and its coarse tuning.} \textbf{a} Normalized pump-band transmission spectrum for a nominal $\mu$OPO device with $H$ = 605~nm, $U$ = 0.33, and $RW$ = 875~nm. Eight TE0 modes ($m_{\rm{p}}$ = 376, ... , 383) are marked with colored circles, and the fitted $Q$ is shown in the right panel. The quoted uncertainty in $Q$ is the one standard deviation value from the fit. \textbf{b} Left: Normalized high-power transmission for mode $m_{\rm{p}} = 380$ of the device from \textbf{a}. We observe the characteristic ``thermal triangle'' when scanning $\lambda_{\rm{p}}$ from blue to red. Different detunings (marked by numerals 1 through 7) generate seven different $\mu$OPOs with $\nu_{\rm{s}}$ differences of approximately one FSR. The corresponding signal spectra are shown in the right panel; here, 0~dB is referenced to 1~mW, i.e., dBm. \textbf{c} Top panel: Compilation of optical spectra generated by the nominal device. The different colors correspond to the $m_{\rm{p}}$ values indicated in \textbf{a}, and bold data correspond to the first OPO spectrum observed when the pump laser is blue-to-red scanned through a given pump mode. The faded data correspond to subsequently observed OPO spectra (e.g., spectra 2-7 in \textbf{b}). Idlers with frequencies below 250 THz are not observed due to the pump/signal waveguide cutoff. Bottom panel: Distribution of the signal and idler frequencies versus simulations. Solid squares are experimental data extracted from the spectra above. Dashed empty squares are estimated based on energy conservation. Solid lines are taken from dispersion simulations.}
\label{Fig3}
\end{figure*} 

\begin{figure*}[t!]
\centering\includegraphics[width=0.9\linewidth]{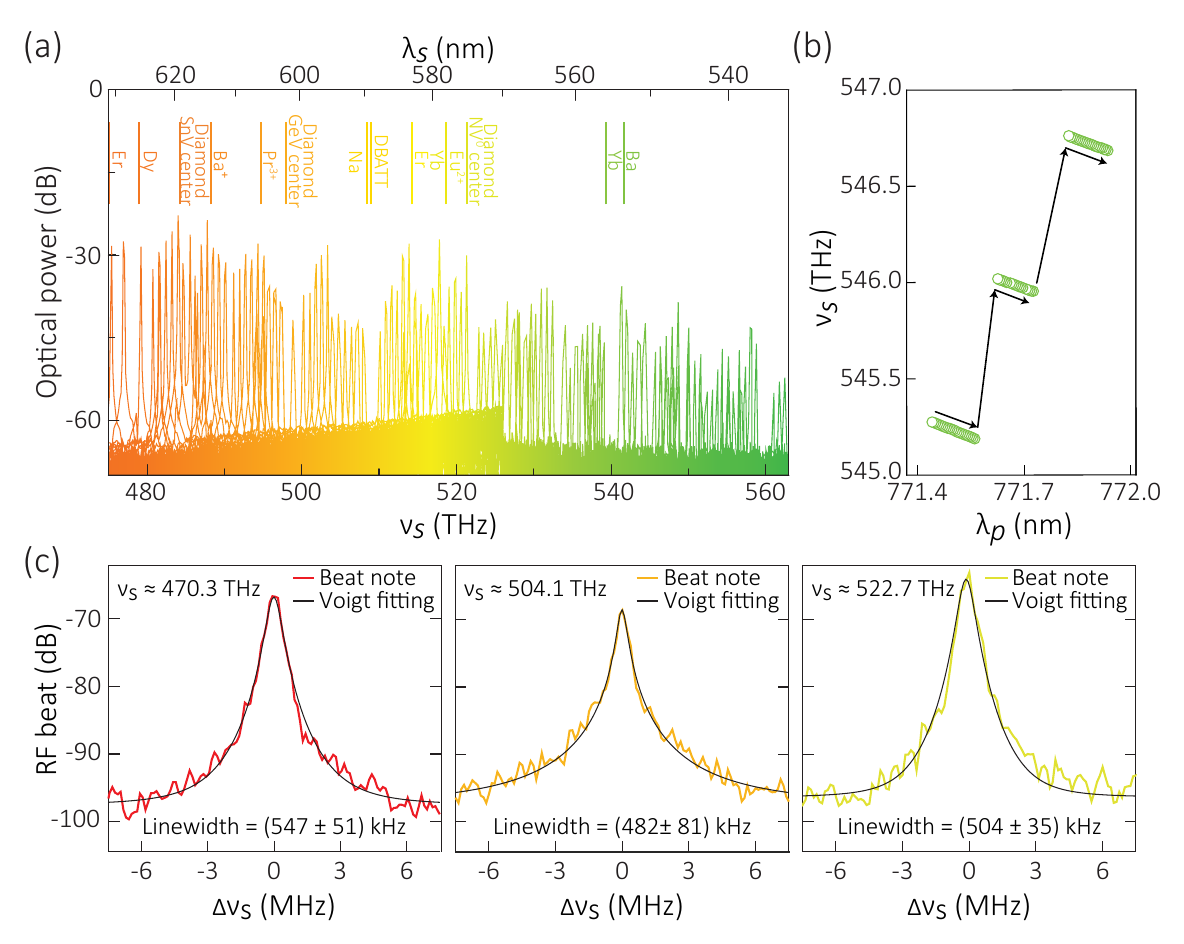}
\caption{\textbf{Continuous fine tuning of the signal frequency ($\nu_{\rm{s}}$) and linewidth measurements.} \textbf{a} Optical spectra compiled from four $\mu$OPO devices with small differences in $RW$. Frequencies greater than $490$~THz are generated using only two devices. The top vertical lines indicate transition wavelengths of various quantum systems within the green gap. \textbf{b} $\nu_{\rm{s}}$ versus pump wavelength in a nominal $\mu$OPO device. Here, the pump wavelength is tuned within a single cavity resonance. Periodic mode hops occur that shift $\nu_{\rm{s}}$ by approximately one FSR. In between mode hops, $\nu_{\rm{s}}$ varies smoothly and continuously with $\nu_{\rm{p}}$, with typical continuous tuning ranges between 50~GHz to 80~GHz. \textbf{c} Radiofrequency heterodyne beat notes between the OPO signal and a Ti:Sapphire sum-frequency-generation (SFG) laser system, along with corresponding fits and fit uncertainties. 0~dB is referenced to 1~mW.}
\label{Fig4}
\end{figure*}

\noindent \textbf{Green gap access.}
Figure \ref{Fig3} depicts the multitude of green gap $\mu$OPOs we make in experiments and illustrates the coarse tuning mechanisms that enable them. In our experiments, we pump TE0 modes in a SiN microring with an amplified external cavity diode laser (ECDL) that is continuously tunable from $765$~nm to $781$~nm. The measurement setup is illustrated in the Appendix (Fig. \ref{figS1}a). Figure \ref{Fig3}a (left panel) depicts the normalized, low-power transmission spectrum of a device with $RR~=~25~\mu$m, $H~=~605$~nm, $RW~=~875$~nm, and $U~=~0.33$. The colored circles mark the eight TE0 modes we can access with the ECDL, and the right panel depicts the normalized transmission zoomed in to $778.761$ nm (wavelength measured with a wavemeter with $\approx$~0.1~pm uncertainty). When we increase the pump power, we can generate OPO, and we observe a characteristic ``thermal triangle'' shape in the transmission spectrum of each resonator mode when $\nu_{\rm{p}}$ is scanned from blue to red detunings, as shown in the left panel of Fig.~\ref{Fig3}b. Moreover, for each specific $m_{\rm{p}}$ value ($380$ in Fig.~\ref{Fig3}b), adjusting the pump-resonator detuning allows us to tune $\nu_{\rm{s}}$ in $FSR$-level increments through a mode-switching mechanism \cite{Stone_PRA_2022}, as shown by the seven different signal spectra presented in the right panel of Fig.~\ref{Fig3}b that correspond to the seven detunings marked in the left panel. The mode switching occurs due to changes in the effective dispersion that arise from Kerr- and thermal-nonlinear mode shifts. Interestingly, we believe that dispersion of the thermo-optic coefficient, $dn/dT$, where $n$ is the refractive index of SiN and $T$ is the modal temperature, must be considered to properly model the mode switching, but such measurements are not found in the existing literature.

To more widely tune $\nu_{\rm{s}}$, we can pump modes with different $m_{\rm{p}}$. In the top panel of Fig.~\ref{Fig3}c, we present optical spectra compiled from a single microring, where each colorband corresponds to a different $m_{\rm{p}}$ value - stepping $m_{\rm{p}}$ by one shifts $\nu_{\rm{s}}$ into a new colorband. Hence, the relatively small $\nu_{\rm{p}}$ tuning range ($\approx8$~THz) enables coarse $\nu_{\rm{s}}$ tuning between $\approx490$~THz and $\approx560$~THz. This coarse tuning mechanism was reported in Ref. \onlinecite{Lu_Optica_2020} and arises from the $\nu_{\rm{p}}$-dependent $\Delta \nu$ spectrum. In the bottom panel, we chart the $\nu_{\rm{s,i}}$ values extracted from the optical spectra above, and we compare our measurements to simulations. We find that simulations accurately predict the $\nu_{\rm{s,i}}$ shifts that result from incrementing $m_{\rm{p}}$. The $\nu_{\rm{p}}$ variation (for a given $m_{\rm{p}}$) that is evident in Fig. \ref{Fig3}c is due to an underlying $RW$ variation (i.e., moving along a line of constant $m_{\rm{p}}$ corresponds to varying $RW$). In Fig.~\ref{Fig3}c, bold data are associated with the first $\mu$OPO observed when $\nu_{\rm{p}}$ is scanned into resonance from blue to red detunings, and faded data are associated with subsequent $\mu$OPOs observed during the scan. We typically observe between six and eight $\mu$OPOs for each $m_{\rm{p}}$ value, as shown in Fig.~\ref{Fig3}b.  

\noindent \textbf{Tunability and coherence.} The spectra in Fig.~\ref{Fig3}c come from one device and exhibit spectral gaps in between colorbands. To address these gaps, we fabricate three more devices with small systematic $RW$ differences. Detailed data for these devices are shown in Fig. \ref{FigS2}a in the Appendix. In Fig. \ref{Fig4}a, we present optical spectra compiled from this set of four devices that address the entire green gap with nearly $FSR$-level resolution. In particular, we use just two devices to generate more than $100$ $\mu$OPOs, including all signal frequencies above $490$~THz, and further dispersion optimization could allow similar performance using only one device. We also note that the highest frequency achieved in our study, as detailed in the Appendix (see Fig. \ref{FigS1}b), is $\approx563.51$~THz, which is nearly $15$~THz beyond the previous record~\cite{Domeneguetti_Optica_2021}. Moreover, since many applications (e.g., spectroscopy of quantum systems - see marked level transitions in Fig.~\ref{Fig4}a) require lasers that are continuously tunable and phase coherent, we proceed to characterize the $\nu_{\rm{s}}$ tunability and measure the $\mu$OPO linewidth. As described above, a useful coupling exists between $\nu_{\rm{p}}$ and $\nu_{\rm{s}}$, where small adjustments to the former (e.g., shifting the pump-resonator detuning) can induce $FSR$-level shifts in the latter. It is also crucial to understand the tuning dynamics in between such mode hops. In Fig.~\ref{Fig4}b, we present measurements of $\nu_{\rm{s}}$, recorded using an optical spectrum analyzer (OSA), as $\nu_{\rm{p}}$ is tuned in our nominal device. We observe that, in between mode hops, the tuning coefficient $d\nu_{\rm{s}}/d\nu_{\rm{p}} \approx 1$ (it is slightly greater than one due to thermo-optic shifts); and typically, we achieve continuous tuning ranges between $50$~GHz and $80$~GHz. Notably, this tuning range depends on the dispersion and can be extended using, e.g., integrated temperature control \cite{moille2022integrated}, and we show some preliminary experimental results with temperature variation in the Appendix (Fig. \ref{figS3}). We note that `bumpiness' in Fig. \ref{Fig4}b (i.e., any deviation from linear tuning) is primarily due to OSA error.

Next, we record radiofrequency spectra from heterodyne beats between the $\mu$OPO signal and a low-noise tunable continuous-wave laser. Here, we use the full-width at half maximum (FWHM) of observed spectral lineshapes to approximate the $\mu$OPO signal linewidth. In particular, we are interested in relative orders of magnitude between the pump laser linewidth ($\approx 300$~kHz) and $\mu$OPO linewidths, and we reserve a more comprehensive noise analysis for future studies. In Fig.~\ref{Fig4}c, we present heterodyne spectra for three $\mu$OPOs with signal frequencies near $470$~THz, $504$~THz, and $523$~THz. Respectively, the fitted lineshapes exhibit FWHM values of $\approx~547$~kHz, $\approx~482$~kHz, and $\approx~504$~kHz, where the figure includes the one standard deviation uncertainties obtained from the fits. These values are commensurate with the pump laser linewidth and demonstrate low added noise from the $\mu$OPO; moreover, they are much smaller than typical III-V diode lasers in this wavelength range. In a more absolute sense, the measured linewidths are already sufficiently small for many high-coherence applications, and future systems could employ injection locking to achieve low-noise operation even with noisy pump lasers \cite{Ling_LaserPhotonicsReviews_2023, Li_Optica_2023, clementi2023chip}.

\section{Discussion}
\noindent In summary, we establish a blueprint, based on widely-separated Kerr OPO, for integrated sources of coherent and highly-tunable light at green-gap frequencies. We generate the most widely-separated $\mu$OPOs to date, with a signal frequency reaching $\approx563.51$~THz and its corresponding idler near $\approx207.28$~THz. Dispersion simulations suggest that cyan emission is possible with our scheme, but it is currently not observed due to parasitic losses in the idler band. Finally, we note that the resonator-waveguide coupling was not optimized to realize large conversion efficiencies, but we expect that conventional strategies to increase coupling (e.g., using pulley waveguide geometries~\cite{Stone_APLPhoton_2022}) will enable efficient green emission. Preliminary measurements, presented in the Appendix, indicate that signal power can be increased $\approx15$~dB by using a separate waveguide for signal extraction, and for a test device we estimate the on-chip signal power to be $\approx 100$~$\mu$W. After further coupling optimization, the on-chip power is expected to exceed $1$~mW \cite{Stone_APLPhoton_2022}.  

\section{Materials and methods}
\noindent Simulations in Fig.~\ref{Fig2} and Fig.~\ref{Fig3} are based on eigenmode calculations of the microrings using the finite-element method. The layout of the devices is prepared using the Nanolithography Toolbox, a free software package provided by the NIST Center for Nanoscale Science and Technology \cite{Balram_JResNatlInst_2018}. A 605-nm-thick layer of SiN is deposited by low-pressure chemical vapor deposition on top of a $3~\mu$m SiO$_2$ layer on a $100$~mm Si wafer. Spectroscopic ellipsometry is employed to measure the layer thicknesses and the wavelength-dependent refractive indices, and the results are fitted using an extended Sellmeier model. The device patterning is realized using positive-tone resist and electron-beam lithography, followed by pattern transfer into the SiN layer through reactive-ion etching using a CF$_4$/CHF$_3$ chemistry. The device undergoes chemical cleaning to remove any residual polymer or resist post-etching and is subsequently annealed at $1100$ $\degree$C in a nitrogen environment for four hours. An oxide lift-off process is executed to ensure air-cladding on the devices while maintaining oxide-cladding over the input and output waveguides. The chip is then diced and polished for lensed-fiber coupling. The microring undercuts are achieved by heated KOH etching at $70$ $\degree$C, with the lateral etching rate estimated by measuring the vertical etching rate and verified through cross-sectional scanning electron microscope images.

\medskip
\noindent \textbf{Data Availability}
The data that supports the plots within this paper and other findings of this study are available from the corresponding author upon reasonable request.

\medskip
\noindent \textbf{Acknowledgements} This work is partially supported by the DARPA LUMOS and NIST-on-a-chip programs. X.L. acknowledges supports from Maryland Innovation Initiative. We thank Dr. Junyeob Song from NIST for his assistance with the focused-ion beam imaging in Fig.~\ref{Fig2}(c). 

\medskip
\noindent \textbf{Author Contributions} X.L., Y.S., and J.S. carried out the device design and simulation. X.L. and Y.S. carried out the fabrication. Y.S., J.S., X.L., and F.Z. carried out the measurements. All authors participated in analysis and discussion of results. J.S. and Y.S. wrote the manuscript with the help from others and K.S. supervised the project.


\noindent \textbf{Competing Financial Interests} The authors declare no competing financial interests.

\appendix
\section{Measurement setup for widely-separated optical parametric oscillation (OPO)}
\begin{figure*}[h!]
\centering\includegraphics[width=0.9\linewidth]{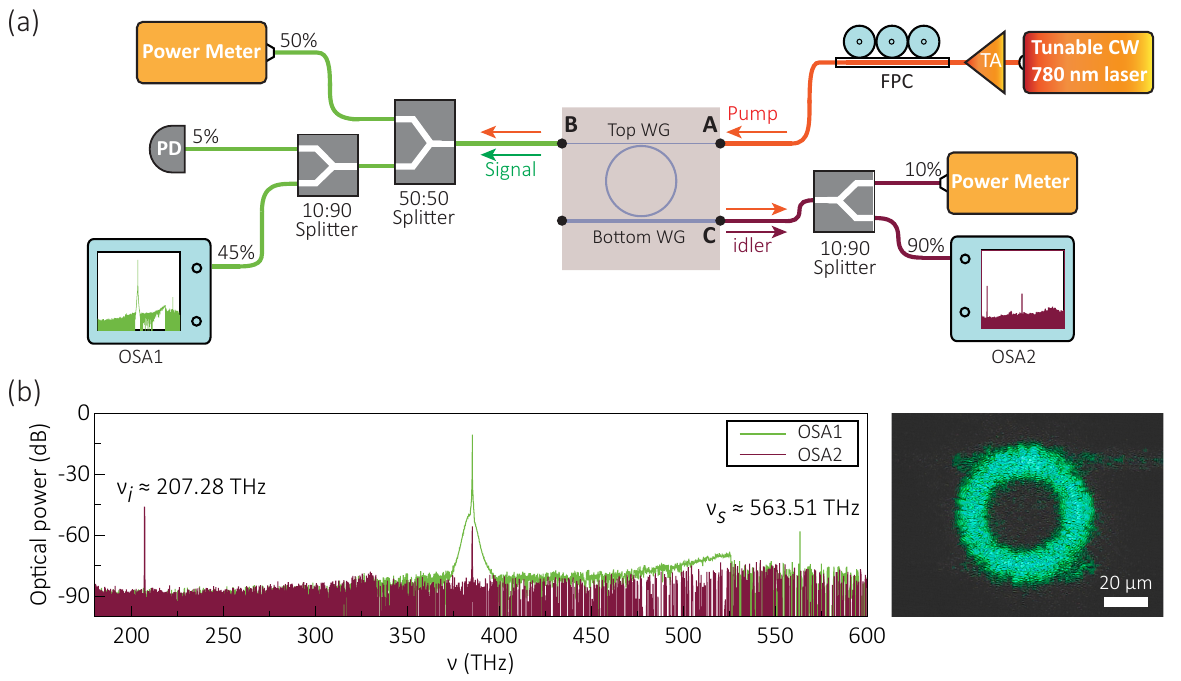}
\caption{\textbf{a} Schematic of our measurement setup. PD: Photodiode, OSA: Optical spectrum analyzer, TA: Tapered amplifier, FPC: Fiber polarization controller. \textbf{b} Left panel: OPO spectrum exhibiting the largest signal-idler separation observed in our study. The green spectrum indicates the signal and pump waves and is captured from the top WG using OSA1. The idler is not visible because it cannot propagate in the top WG. The dark red spectrum indicates the idler and pump waves and is captured from the bottom WG using OSA2. The signal is significantly undercoupled to the bottom WG; hence, it is not detectable. Right panel: Optical microscope image of the microring during OPO. Scattered signal light (green) is clearly visible.}
\label{FigS1}
\end{figure*}

\noindent Figure~\ref{FigS1}a illustrates our measurement setup for OPO experiments. We use lensed fibers to in/out-couple light to/from inverse tapered waveguides (WGs). The estimated insertion loss for the pump is $\approx3$ dB per facet. The top WG is employed for in- and out-coupling the pump and signal waves, while the idler is extracted from the bottom WG. We use this configuration for both linear cavity transmission measurements (as shown in Fig.~3a of the main text) and high-power OPO measurements. For linear cavity transmission measurements, we use a continuous-wave (CW) laser whose wavelength is tunable from $765$~nm $781$~nm. We attenuate the laser power to sub-microWatt levels and adjust its polarization to overlap the microring TE modes. In the case of OPO measurements, we boost the pump laser power to $\approx300$~mW using a tapered amplifier. Spectra are obtained with two optical spectrum analyzers (OSAs). OSA1 collects light from the top WG and OSA2 collects light from the bottom WG, as shown in Fig. \ref{FigS1}a. To illustrate the need for two WGs, in Fig.~\ref{FigS1}b we present the OPO spectrum with the widest signal-idler separation observed in our study. The signal frequency is $\nu_{\rm{s}}\approx$~$563.51$~THz, which corresponds to an idler frequency $\nu_{\rm{i}}\approx$~$207.28$~THz. In the spectrum extracted from the top WG, we detect (do not detect) measurable amounts of signal (idler) power, and vice versa. The right panel displays a microscope image of the microring in which scattered signal light is clearly visible.

\section{Kerr OPO covering the green gap}
\begin{figure*}[ht]
\centering\includegraphics[width=0.9\linewidth]{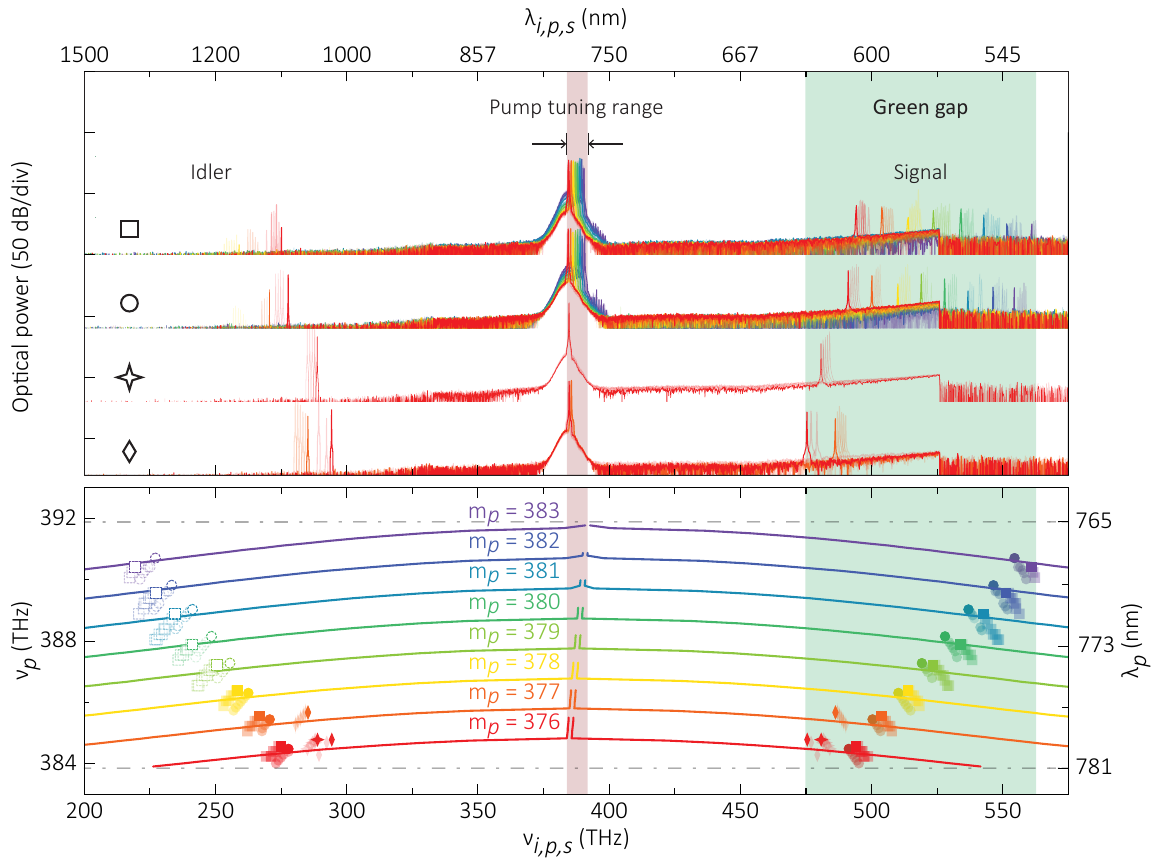}
\caption{Top panel: Compilation of optical spectra generated from four devices. The different colors correspond to the $m_{\rm{p}}$ values as marked in the bottom panel. The bold data correspond to the first OPO spectrum observed when the pump laser is blue-to-red scanned through a given pump mode. The faded data correspond to subsequently observed OPO spectra (the same as in Fig. 3c in the main text). Idlers with frequencies below 250 THz are not observed due to the pump/signal waveguide cutoff. Bottom panel: Distribution of the signal and idler frequencies from these four devices as extracted from the spectra above. Solid lines are taken from dispersion simulations.}
\label{FigS2}
\end{figure*}

In Fig. 4a of the main text, we show a set of overlaid optical spectra compiled from four devices. Here, we separate the spectra and sort them according to which device was used to generate them; see Fig.~\ref{FigS2}. We note, in the two devices indicated by square and circle symbols, OPO spectra via pumping of eight different pump modes, covering the spectral range from $\approx490$~THz to $\approx560$~THz with $>100$ total spectra. For devices marked with a star or diamond, we needed only to pump one or two modes to target the remaining portion of the green gap, from $\approx475$~THz to $\approx490$~THz. Therefore, the total span of signal wavelengths accessible using these four devices is much greater than we present, though here we obviously focus on the green gap.

\section{Continuous tuning with temperature variation}
\begin{figure*}[ht]
\centering\includegraphics[width=0.9\linewidth]{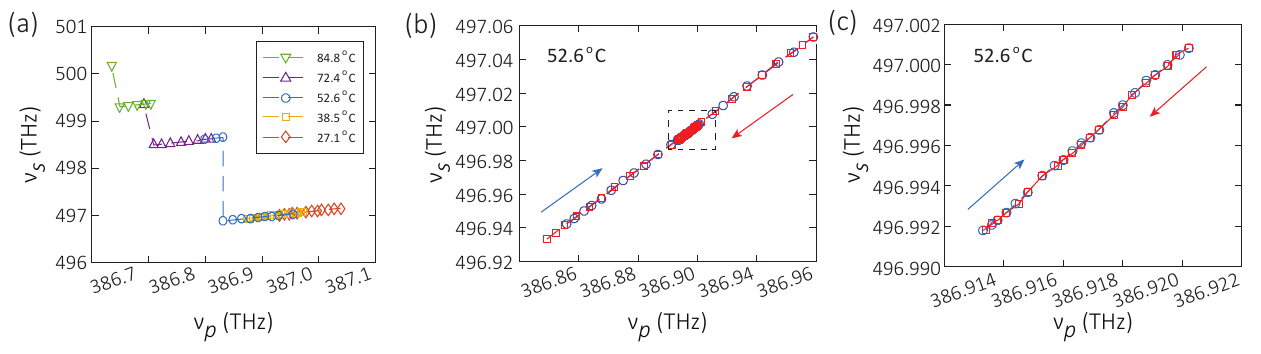}
\caption{\textbf{a} Wavemeter measurements of the OPO signal frequency, $\nu_{\rm{s}}$, versus pump frequency, $\nu_{\rm{p}}$, for different temperature setpoints in a nominal device. \textbf{b} Higher-resolution measurement of $\nu_{\rm{s}}$ versus $\nu_{\rm{p}}$ at a temperature setpoint of 52.6 degrees Celsius. \textbf{c} Same measurement zoomed into the region indicated by the dashed box in (b).}

\label{FigS3}
\end{figure*}

In the main text, we demonstrate continuous $\nu_{\rm{s}}$ tuning up to $80$~GHz using $\nu_{\rm{p}}$ actuation. Here, we show one route to extend the tuning range using temperature variation. We change the temperature of our sample by heating (using a resistive strip heater) the metal mount on which it sits. In Fig. \ref{FigS3}a, we present measurements of $\nu_{\rm{s}}$ (recorded using a wavemeter) versus $\nu_{\rm{p}}$ for different temperature setpoints ranging from $27$ degrees to $85$ degrees Celsius. As $\nu_{\rm{p}}$ is tuned, we observe continuous $\nu_{\rm{s}}$ tuning with intermittent mode hops, as described in the main text. However, for different temperatures, the structure of these mode hops is altered, increasing the range of accessible $\nu_{\rm{s}}$ values to upwards of $200$ GHz in some cases. In Figs. \ref{FigS3}b-c, we have zoomed into a narrower range of $\nu_{\rm{s}}$ values to establish that our tuning is truly continuous and reversible.


\maketitle
\section{Improving signal extraction efficiency}

\begin{figure*}[ht!]
\centering\includegraphics[width=0.9\linewidth]{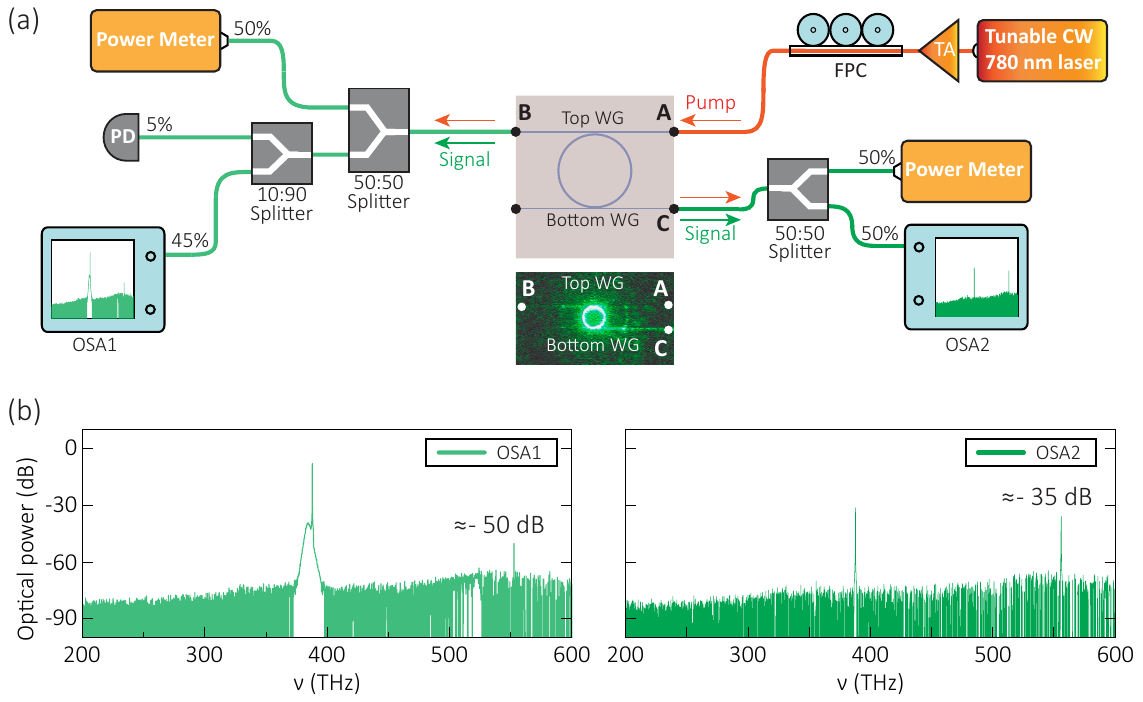}
\caption{\textbf{a} Measurement setup for OPO devices using a separate bottom WG for signal extraction. \textbf{b} OPO spectrum recorded using both OSA1 (left panel) and OSA2 (right panel). The optical power extraction is $\approx15$~dB more efficient in the bottom WG than the top WG. Here, 0 dB is referenced to 1 mW, i.e., dBm. After calibration of losses, the on-chip signal power in the bottom WG is estimated to be $\approx$~100~$\mu$W.}
\label{FigS4}
\end{figure*}

For devices studied in the main text, we use a top WG (see Fig. \ref{FigS1}a) coupling gap of $100$~nm, which corresponds to near-critical waveguide-resonator coupling at the pump wavelength, but necessarily leaves the OPO signal undercoupled. Such undercoupling is responsible for the relatively small amounts of extracted signal power measured in the main text. To prove that greater green-gap extraction efficiencies are possible, we here demonstrate the possibility to use a separate WG for out-coupling signal light. As shown in Fig.~\ref{FigS4}a, we introduce a bottom WG with a coupling gap of $70$~nm that is designed to be cut-off for pump wavelengths and near critically coupled for signal wavelengths (instead of the idler, as in our previous experiments). In this configuration, OSA1 collects light from the top WG and OSA2 collects light from the bottom WG. The optical microscope image clearly shows that signal light extracted to the botton WG is brighter than the top WG. Indeed, OSA measurements confirm that the bottom WG is more than 10x better at extracting signal light from the microring than the top WG, as shown in Fig. \ref{FigS4}b. Furthermore, we calibrate the optical losses during propagation from the bottom WG to OSA2 to be $\approx 25$~dB. Hence, we estimate the on-chip signal power for this OPO device to be $\approx100$~$\mu$W. Such powers are already sufficient for many applications and demonstrate, in principle, that optimized coupling configurations greatly improve efficiency. In the future, we expect further coupling optimization (e.g., through use of pulley geometries) to enable milliWatt-level on-chip signal powers.

\bibliography{kOPO}

\end{document}